\documentclass[prl,a4paper,superscriptaddress,twocolumn,amsmath,amssymb]{revtex4}

\usepackage{graphicx}
\usepackage{bm}
\graphicspath{{./Figs/}}
\usepackage{dsfont}
\usepackage{epstopdf}
\usepackage{hyperref}

\newcommand {\figwidth} {3.4in}

\newcommand{\beq}{\begin{equation}}
\newcommand{\eeq}{\end{equation}}

\newcommand{\sitwonine}{$^{29}$Si}
\newcommand{\sitwoeight}{$^{28}$Si}
\newcommand{\biiso}{$^{209}$Bi}

\newcommand{\beqa}{\begin{eqnarray}}
\newcommand{\eeqa}{\end{eqnarray}}

\newcommand{\tonee}{$T_{\rm{1e}}$}

\newcommand{\ttwoe}{$T_{\rm{2e}}$}

\begin{document}

\title{Decoherence mechanisms of \biiso~donor electron spins in isotopically pure \sitwoeight}

\author{Gary Wolfowicz}
\email{gary.wolfowicz@materials.ox.ac.uk}
\author{Stephanie Simmons}
\affiliation{London Centre for Nanotechnology, University College London, London WC1H 0AH, UK} 
\affiliation{Dept.\ of Materials, Oxford University, Oxford OX1 3PH, UK} 

\author{Alexei M. Tyryshkin}
\affiliation{Dept.\ of Electrical Engineering, Princeton University, Princeton, New Jersey 08544, USA}

\author{Richard E. George}
\affiliation{Dept.\ of Materials, Oxford University, Oxford OX1 3PH, UK} 

\author{Helge Riemann}
\author{Nikolai V. Abrosimov}
\affiliation{Institute for Crystal Growth, Max-Born Strasse 2, D-12489 Berlin, Germany}

\author{Peter Becker}
\affiliation{Physikalisch-Technische Bundesanstalt, D-38116 Braunschweig, Germany}

\author{Hans-Joachim Pohl}
\affiliation{Vitcon Projectconsult GmbH, 07745 Jena, Germany}

\author{Stephen A. Lyon}
\affiliation{Dept.\ of Electrical Engineering, Princeton University, Princeton, New Jersey 08544, USA}

\author{Mike L. W. Thewalt}
\affiliation{Dept.\ of Physics, Simon Fraser University, Burnaby, British Columbia V5A 1S6, Canada}

\author{John~J.~L.~Morton}
\email{jjl.morton@ucl.ac.uk}
\affiliation{London Centre for Nanotechnology, University College London, London WC1H 0AH, UK} 
\affiliation{Dept.\ of Electronic \& Electrical Engineering, University College London, London WC1E 7JE, UK} 

\date{\today}

\begin{abstract}
Bismuth ($^{209}$Bi) is the deepest Group V donor in silicon and possesses the most extreme characteristics such as a 9/2 nuclear spin and a 1.5~GHz hyperfine coupling. These lead to several potential advantages for a Si:Bi donor electron spin qubit compared to the more common phosphorus donor. Previous studies on Si:Bi have been performed using natural silicon where linewidths and electron spin coherence times are limited by the presence of $^{29}$Si~impurities.  Here we describe electron spin resonance (ESR) and electron nuclear double resonance (ENDOR) studies on $^{209}$Bi~in isotopically pure $^{28}$Si. ESR and ENDOR linewidths, transition probabilities and coherence times are understood in terms of the spin Hamiltonian parameters showing a dependence on field and $m_I$ of the $^{209}$Bi~nuclear spin.
We explore various decoherence mechanisms applicable to the donor electron spin, measuring coherence times up to 700~ms at 1.7~K at X-band, comparable with $^{28}$Si:P. The coherence times we measure follow closely the calculated field-sensitivity of the transition frequency, providing a strong motivation to explore `clock' transitions where coherence lifetimes could be further enhanced.
\end{abstract}

\maketitle

Amongst the first proposals for quantum information processing (QIP) in solid state devices was that by Kane in 1998, using phosphorus (P) dopants in silicon, the first of the group V donors~\cite{Kane1998}. 
The choice of the P-dopant in particular can be attributed to its use in classical silicon technology as donors, in addition to the simplicity of the spin system which consists of an electron and nuclear spin 1/2.
Si:P has been extensively studied as a potential qubit~\cite{Tyryshkin2011a, Morton2011} and is more than ever one of the leading candidates in the quest for a solid-state quantum computer \cite{Morello2010, Fuechsle2012}. 
It begs the question, however, of whether other donors of the same group, i.e.\ arsenic (As), antimony (Sb) and bismuth (Bi), could have the same if not better properties than P. As the deepest group V donor, Bi has the largest nuclear spin (9/2) and the largest hyperfine coupling (1.4754~GHz) \cite{Feher1959}, and has received increasing attention over the past two years~\cite{Sekiguchi2010, George2010, Morley2010a, Belli2011, Weis2012a}. A single donor has a 20 dimensional Hilbert space, allowing either more storage space or a more robust encoding of information. The large hyperfine coupling enables shorter nuclear manipulation times and a zero field splitting of 7.3~GHz, making Si:Bi useful as memory for hybrid superconducting circuits~\cite{George2010, Kubo2011, Schuster2010}. At low magnetic fields, the spin Hamiltonian provides so-called `clock' transitions for the electron spin, where the transition frequency between two states is insensitive to first order in magnetic field fluctuations of the environment~\cite{Mohammady2010, Mohammady2010a, Balian2012}. On the other hand, high vapour pressure and low solubility of the donor in silicon complicates the doping process during growth \cite{Riemann2006}, while the high atomic weight (209) increases defect density during ion implantation \cite{Weis2012a}.

Previous works \cite{George2010,Morley2010a,Belli2011,Sekiguchi2010} on bismuth were realized using doped natural silicon crystals where the presence of other silicon isotopes, in particular \sitwonine~with a nuclear spin of $1/2$, resulted in spectral broadening and coherence times limited to about 1~ms. We investigate here isotopically pure \sitwoeight~crystals~\cite{Becker2010}, where the Bi was introduced during crystal growth using the method developed earlier for natural Si~\cite{Riemann2006}. 
The residual concentration of \sitwonine~is 50~ppm, allowing us to explore the intrinsic properties of the bismuth donor electron spin in silicon. Pulsed electron spin resonance (ESR) experiments were performed using a X-band (9.74~GHz) Bruker Elexsys E680 and E580 spectrometers. Unless otherwise mentioned, the concentration of activated Bi is $2\times10^{15}$~cm$^{-3}$.

\begin{figure}[!ht] 
\centerline{\includegraphics[width=\figwidth]{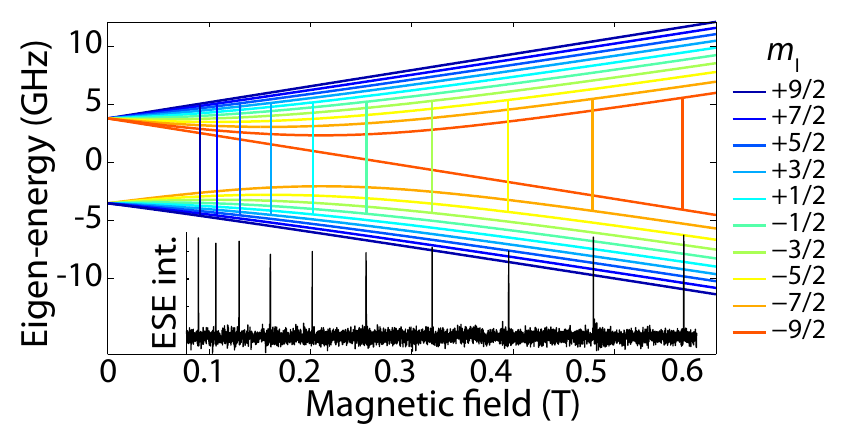}}\caption{Eigenenergies of Si:Bi in frequency units. Twenty eigenstates arise (for $B_0>0$) from the electron spin $S =1/2$ and the nuclear spin $I = 9/2$. Eigenstates are coloured according to their $m_I$ value in limit of large $B_0$, and allowed transitions at X-band (9.7 GHz) are highlighted. Inset shows the corresponding X-band electron spin echo-detected spectrum.} 
\label{SiBiSystem}
\end{figure}

\begin{figure}[!ht] 
\centerline{\includegraphics[width=\figwidth]{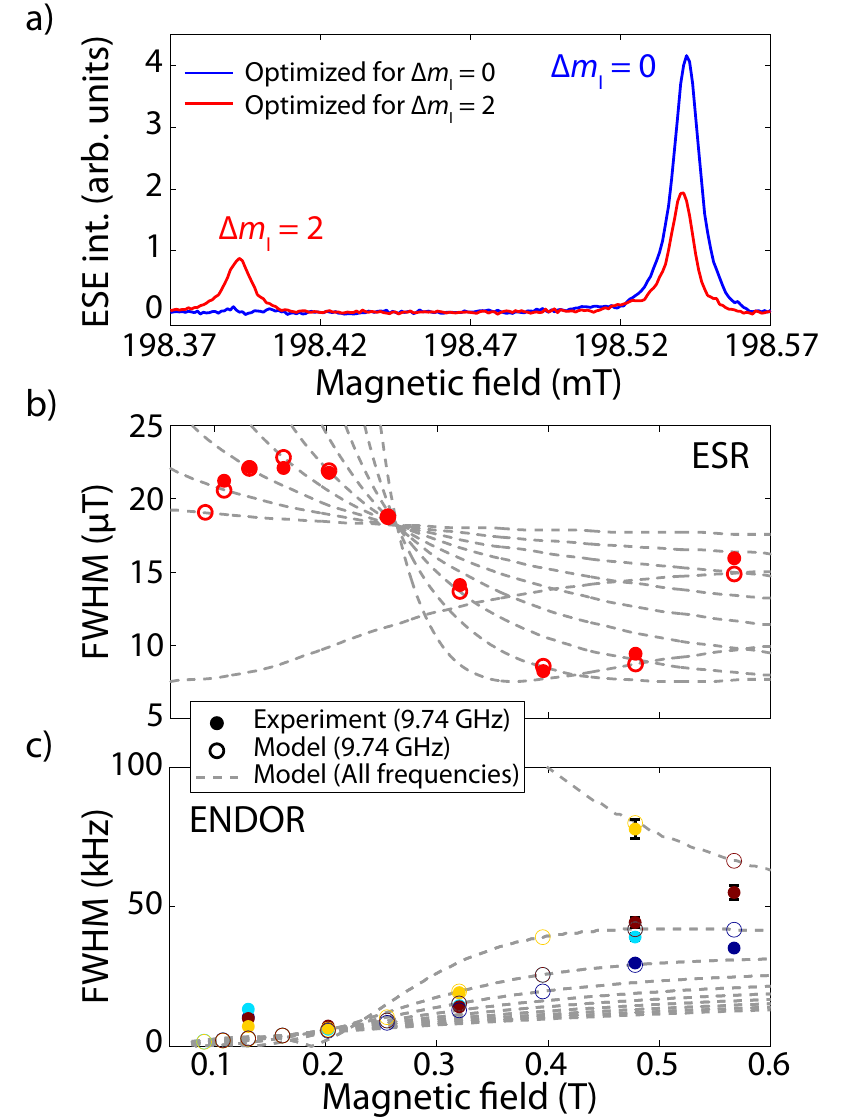}}\caption{a) Typical allowed ($\Delta m_I=0$) and forbidden transitions ($|\Delta m_I|=2$) observed by ESE-detected field sweep. The microwave power is optimized separately for each transition. b,c) Simulated ESR and ENDOR linewidths in light grey are calculated assuming a Gaussian spread in $A$ and $B_0$ of about 60~kHz and 3~$\mu$T respectively. Empty points (model) and filled points (experiment) correspond to the transitions at 9.74~GHz. 
ESR and ENDOR linewidths were measured at 25~K and 10~K respectively.
For the ENDOR linewidths, the colours are guides to the eye to facilitate comparison of the experimental and simulated values.
At the lowest measured field, the ENDOR linewidths only provide an upper bound due to experimental limitations.}
\label{linewidths}
\end{figure}

The bismuth donor coupled nuclear/electron spin system is described in frequency units by an isotropic spin Hamiltonian:
\begin{equation}
H_0 = B_0 (\gamma_E S_z \otimes \mathds{1} + \gamma_N \mathds{1} \otimes I_z) + A\vec{S}.\vec{I}
\label{eq:Hamiltonian}
\end{equation}

where the two first terms correspond to the electronic ($S$) and nuclear ($I$) Zeeman interactions with an external field $B_0$ and the last term corresponds to the hyperfine coupling $A$ (Fig.~\ref{SiBiSystem}). In addition to the normal ESR-allowed transitions $[\Delta m_S = \pm 1, \Delta m_I =0]$ (high field notation), eight `forbidden' transitions $[\Delta m_S = \pm 1, \Delta m_I = \mp 2]$ can be found at the lower field side of the ESR lines for $-7/2 \leq m_I \leq +7/2$ (Figure \ref{linewidths}a). The separation between allowed and forbidden peaks increases from $-30$~$\mu$T for $m_I = +7/2$ to $-270$~$\mu$T for $m_I = -7/2$. Simulations using EasySpin~\cite{Stoll2006} confirm the non-negligible intensity of these forbidden transitions, especially for low $|m_I|$ values which are the most mixed in terms of $m_S$ and $m_I$. An electron spin echo (ESE) intensity ratio between the allowed/forbidden transitions of nearly 0.25 was measured after optimization of the microwave power to each. Therefore, these forbidden transitions offer an additional way to manipulate more states of the 20 dimensional Hilbert space at the speed of a typical ESR pulse (tens of ns).

These $|\Delta m_I| = 2$ transitions were not observed in earlier studies using $^{\rm nat}$Si:Bi because unresolved hyperfine coupling to the surrounding \sitwonine~broadens the ESR lines to about 0.4~mT. In contrast, the continuous wave (CW)-ESR linewidths in $^{28}$Si:Bi vary between 8.2~$\mu$T and 22~$\mu$T (Figure \ref{linewidths}b). Their $m_I$ dependence can be explained by a full width half maximum (FWHM) hyperfine distribution between 40 and 90~kHz (depending on the sample), and a static $B_0$ field inhomogeneity about 3~$\mu$T (a typical value for ESR). The hyperfine spread corresponds to about 30~ppm of the total value and arises most likely from lattice strains (the crystal was found to contain dislocations). Figure \ref{linewidths}c shows electron-nuclear double resonance (ENDOR) linewidths measured for about half of the 36 transitions. These can be explained by the same distribution in the hyperfine coupling $A$ (the $B_0$ inhomogeneity has negligible influence on the ENDOR linewidth).

The electron spin coherence time \ttwoe\ was measured using a Hahn echo to eliminate the effect of static inhomogeneities. Using this method, a \ttwoe\ of 20~ms can be measured at 5~K, however this technique adds its own type of decoherence to the system known as instantaneous diffusion (ID). The microwave $\pi$-pulse in the $\pi/2-\pi-echo$ sequence flips all the spins in the sample, creating a change in the local magnetic field experienced by each spin. This additional dephasing mechanism, purely experimental, can be reduced by shortening the $\pi$-flip to a smaller tip-angle rotation, at the cost of less signal/noise. By extrapolation to a tip angle of zero, the intrinsic $T_{\rm 2e,int}$ can be found \cite{Schweiger2001}:

\begin{equation}
\frac{1}{T_{\rm 2e}} = \frac{1}{T_{\rm 2e,int}} + C\cdot(2\pi\gamma_{\rm eff})^2 \cdot\frac{\pi}{9\sqrt{3}} \mu_0 \hbar \cdot \sin \left(\frac{\theta}{2}\right)^2
\label{eq:ID}
\end{equation}

where $\theta$ is the tip-angle of the second microwave pulse and $C$ is the effective spin concentration (1/10$^{\rm th}$ of the actual donor concentration as only one out of the 10 ESR lines is excited by a given pulse and participates in ID). $\gamma_{\rm eff}$ is the effective gyromagnetic ratio, equal to $df/dB$, the gradient of the transition frequency with respect to the magnetic field.  It represents both the sensitivity of the spin to the environment and its magnetic dipole strength~\footnote{A second effect of small tip-angles is to prevent decoherence from so-called `direct' flip-flops between a central Bi spin and one of its neighbours (as the probability of flipping both spins is low, the applied microwave pulse refocuses the flip-flop).}. 

\begin{figure}[t] 
\centerline{\includegraphics[width=\figwidth]{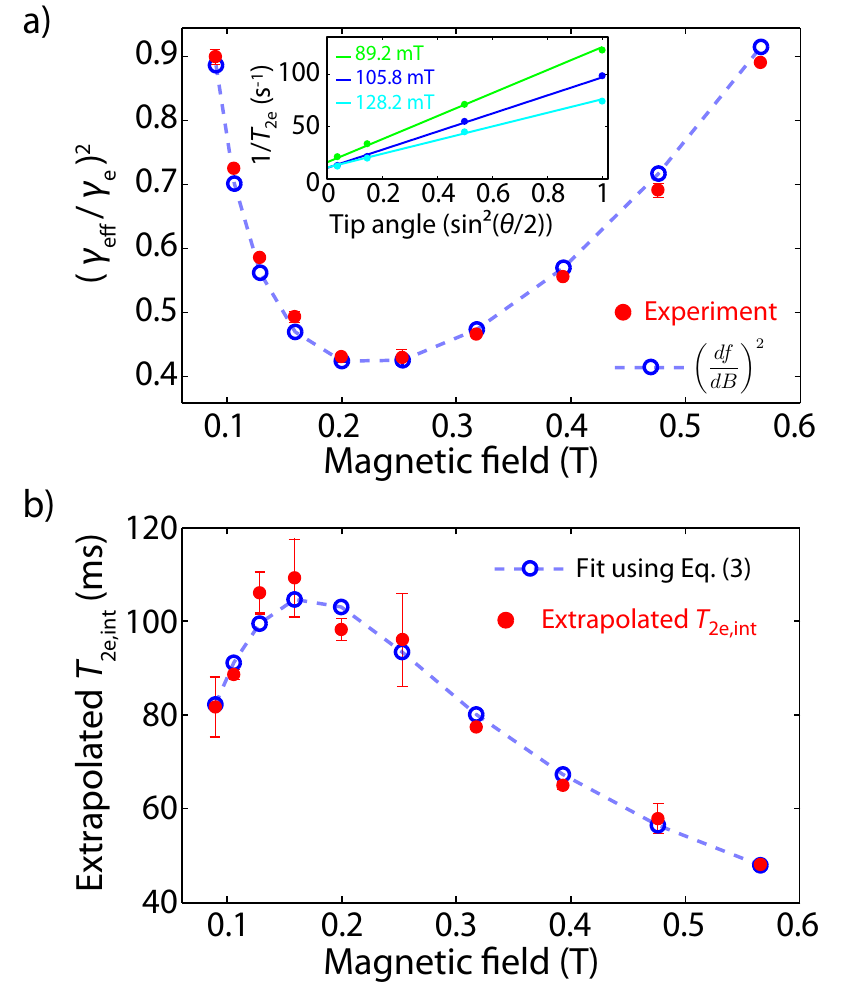}}\caption{Measurements of $T_{\rm 2e,int}$ for the ten ESR transitions in Si:Bi. a)~
ID slopes resulting from the fit of \ttwoe\ times obtained using $\pi, \pi/2, \pi/4$ and $\pi/8$ pulses  as measured at 5~K for each $m_I$ (some examples are shown in inset). The slopes vary according to $(df/dB)^2$, as expected from dipole-dipole interactions between neighbouring Bi electron spins. b) Extrapolated $T_{\rm 2e,int}$~from the fit in (a) in the limit where $\theta \rightarrow 0$. The model shown in blue takes into account various sources of fluctuating magnetic field.} 
\label{T2eField}
\end{figure}
\begin{figure}[t] 
\centerline{\includegraphics[width=\figwidth]{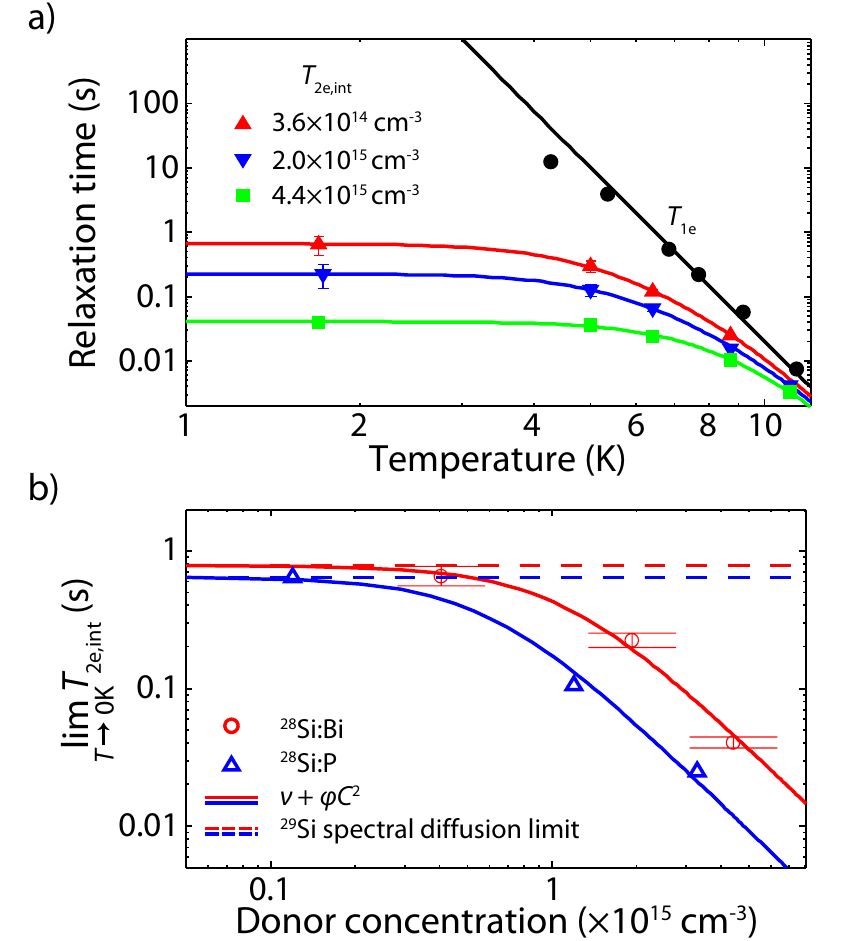}}\caption{Measurements of extrapolated $T_{2e,int}$ as a function of temperature and donor concentration. a) Temperature dependence of  $T_{2e,int}$~and \tonee\ for different donor concentrations. As temperature is lowered, $T_{2e,int}$ is limited first by \tonee\ directly, then by spin flips (\tonee) of neighbouring Bi electron spins, and finally by flip-flops of the neighbours. \tonee\ was measured using saturation or inversion recovery. Around 6 and 8~K, stretched exponential decays were observed as spin flips (\tonee) of neighbours becomes the dominant mechanism. In these cases, a better measure of the coherence is the ``memory time'', shown as \ttwoe\ here, when the curve's amplitude reaches 1/e. b) Fitted decay times in the limit where $T\rightarrow0$ are shown from (a) for \sitwoeight:Bi, and from Ref \cite{Tyryshkin2011a} for \sitwoeight:P. The donor concentration dependence is expected to be quadratic for flip-flops until the effect of residual \sitwonine~nuclear spins becomes dominant at very low donor concentration. } 
\label{T2eTC}
\end{figure}

For each sample with a given Bi donor concentration, we measured \ttwoe\ as a function of tip-angle for each of the different $m_I$ (inset in Fig.~\ref{T2eField}a), and fit the results to Eq.\ (\ref{eq:ID}) with a single concentration $C$ and $\gamma_{\rm eff}$ for each $m_I$ as fitting parameters. In all cases, the best-fit value of $C$ matched that extracted from electrical resistivity measurements to within 10$\%$. The fitted values for $\gamma_{\rm eff}$ are shown in Fig.~\ref{T2eField}a and are in excellent agreement with the theoretical $df/dB$ for each transition, as calculated from the spin Hamiltonian in Eq. (1). In addition to the slope, the intercepts at zero tip-angle were also extracted from the fit providing the extrapolated $T_{\rm 2e,int}$ which are summarized in Figs. \ref{T2eField} and \ref{T2eTC} as a function of $m_I$, temperature and donor concentration. 

The different values for $df/dB$ for the ten ESR lines of Si:Bi offer an ideal opportunity to explore the contributions of decoherence mechanisms for donor electron spins in silicon, as studied previously in \sitwoeight:P \cite{Tyryshkin2011a}. \ttwoe\ is limited directly by \tonee\ above about 9~K, while below this temperature decoherence is driven by spin flips (\tonee) or energy-conserving flip-flops of neighbouring donor electron spins. 
Under both of these mechanisms, the effect on decoherence of a central spin is the product of both the sensitivity $df/dB$ of the central spin and the effective fluctuating magnetic field produced by neighbouring spins. This effective field is related to the magnetic dipole strength of the neighbours, $df/dB$, however, since each neighbouring donor can be in any one of ten $m_I$ states, the relevant $df/dB$ value must be averaged over different $m_I$ at the particular value of the applied magnetic field $B_0$. Therefore, the effect of (\tonee) spin flips of neighbours can be simply modelled by an average dipole strength $\langle df/dB \rangle$. On the other hand, the effect of flip-flops between neighbours is the average product of the flip-flop rate and the magnetic field strength. At X-band, calculations show that this contribution has a similar trend to \tonee~of neighbours. In summary, the overall decoherence rate can be expressed as:
\begin{equation}
1/T_{\rm 2e,int} = \frac{df}{dB}\times \left( \alpha + \beta \left\langle \frac{df}{dB} \right\rangle \right)
\label{eq:T2evsVB}
\end{equation}
where $\alpha$ is an additional field fluctuation, possibly from the external magnet, and $\beta$ represents both neighbour-\tonee\ and flip-flop mechanisms as they are indistinguishable here. This is in excellent agreement with the observed experimental $m_I$ dependence of $T_{\rm 2e,int}$, shown in Fig. \ref{T2eField}b at 5~K.

The temperature dependence of $T_{\rm 2e,int}$ in Fig.\ \ref{T2eTC}a arises from the combination of the  (temperature independent) neighbour flip-flop process and the (strongly temperature dependent) \tonee. The latter appears twice in the sum of rates which determine \ttwoe: directly as \tonee\ for spin flips of the central spin and as $\sqrt{T_{\rm {1e}}}$ for spin flips of the neighbours~\cite{Tyryshkin2011a}. \tonee\ is independent of concentration for small donor concentrations \cite{Feher1959a}. As temperature decreases, \tonee~increases above seconds and neighbour flip-flops become the limiting factor for coherence. 

By fitting the temperature dependence of \ttwoe\ according to the mechanisms above, we extract a low-temperature limit of \ttwoe\ corresponding to the neighbour flip-flop mechanism, and plot this time in Fig.~\ref{T2eTC}b against concentration for the three samples we have studied here. For comparison, \sitwoeight:P values from Ref \cite{Tyryshkin2011a} under similar experimental conditions are also shown.

For concentrations above a few 10$^{14}$~cm$^{-3}$, the effect of flip-flops play a limiting role in \ttwoe, which follows a quadratic dependence on concentration~\cite{DeSousa2003a}. This dependence can be understood by considering that concentration affects both the flip-flop rate, and the sensitivity of the central spin to a neighbour flip-flop. At lower concentrations, the effect of residual \sitwonine~impurities becomes non-negligible and limits the coherence time to about 2~s. The difference in sensitivity to \sitwonine~spectral diffusion between Bi and P can be accounted for, within experimental errors, by the difference in $df/dB$ (a factor of 0.65). 

The average dipole coupling rate between two nearest neighbour Bi donor electron spins is about 10~Hz for a donor concentration of 10$^{15}$~cm$^{-3}$. In contrast, the ESR linewidth is due to inhomogeneity in both $B_0$ and $A$ corresponding to broadening of about 60 and 200~kHz, respectively. Both of these linewidths are much greater than the dipole coupling rate and would largely suppress any nearest neighbour flip-flops if they arose from inhomogeneity across a length-scale comparable to donor-donor separation. For this reasons we conclude that observed spread in hyperfine coupling strength $A$ derives from long-distance variations during the crystal growth process, consistent with strain fields from dislocations.


The temperature dependence of \tonee\ for Si:Bi was studied by Belli \textit{et al.}~\cite{Belli2011} and found to be driven by an Orbach process above 25~K. Below this temperature, a first-order $T^{-7}$ Raman process has been suggested in recent papers \cite{George2010, Morley2010a, Belli2011}, however we find that a $T^{-9}$ Raman process from the spin-orbit interaction to be a better fit. The former process ($T^{-7}$) is expected to be frequency-dependent (\tonee\ $\propto f^{-2}$) while the latter ($T^{-9}$) has no frequency dependence~\cite{Jr1963}. We measured \tonee~as a function of temperature at 7~GHz and found the same values as at 9.74~GHz, following $T^{-9}$ down to 5~K. 

We have measured a \ttwoe\ time of Bi donors in \sitwoeight\ of up to 700~ms at 1.7~K (using a refocussing pulse of angle 0.08$\pi$ radians), comparable with that for P donors. Throughout this study of the various decoherence mechanisms present in \sitwoeight:Bi we have observed that the electron spin decoherence rates follows $df/dB$ to at least first order. This provides a motivation to study \ttwoe\ at various `clock' transitions present in the range 5.2--7.3~GHz, at which $df/dB$ drops to zero. At these points, we would expect \ttwoe\ to have a greatly reduced sensitivity to donor or \sitwonine~concentration, and approach the fundamental limit of \tonee\ which exceeds one second below 6~K. 


This research is supported by the EPSRC through the Materials World Network (EP/I035536/1), CAESR (EP/ D048559/1) and a DTA, as well as by the European Research Council under the European CommunityÕs Seventh Framework Programme (FP7/2007-2013) / ERC grant agreement no. 279781. 
Work at Princeton was supported by NSF through Materials World Network (DMR-1107606) and through the Princeton MRSEC (DMR-0819860), and also by NSA/LPS through LBNL (100000080295). S.S. is supported by the Violette and Samuel Glasstone Fund and St. John's College Oxford. J.J.L.M. is supported by the Royal Society. 

\bibliography{library}

\begin{thebibliography}{10}

\bibitem{Kane1998}
B.~E. Kane, Nature {\bf 393},  133  (1998).

\bibitem{Tyryshkin2011a}
A.~M. Tyryshkin {\it et~al.}, Nature Materials {\bf 11},  143  (2012).

\bibitem{Morton2011}
J.~J.~L. Morton, D.~R. McCamey, M.~A. Eriksson, and S.~A. Lyon, Nature {\bf
  479},  345  (2011).

\bibitem{Morello2010}
A. Morello {\it et~al.}, Nature {\bf 467},  687  (2010).

\bibitem{Fuechsle2012}
M. Fuechsle {\it et~al.}, Nature Nanotechnology {\bf 7},  242  (2012).

\bibitem{Feher1959}
G. Feher, Phys. Rev. {\bf 114},  1219  (1959).

\bibitem{Sekiguchi2010}
T. Sekiguchi {\it et~al.}, Phys. Rev. Lett. {\bf 104},  137402  (2010).

\bibitem{George2010}
R.~E. George {\it et~al.}, Phys. Rev. Lett. {\bf 105},  67601  (2010).

\bibitem{Morley2010a}
G.~W. Morley {\it et~al.}, Nature Materials {\bf 9},  725  (2010).

\bibitem{Belli2011}
M. Belli, M. Fanciulli, and N.~V. Abrosimov, Phys. Rev. B {\bf 83},  235204
  (2011).

\bibitem{Weis2012a}
C.~D. Weis {\it et~al.}, App. Phys. Lett. {\bf 100},  172104  (2012).

\bibitem{Kubo2011}
Y. Kubo {\it et~al.}, Phys. Rev. Lett. {\bf 107},  220501  (2011).

\bibitem{Schuster2010}
D. Schuster {\it et~al.}, Phys. Rev. Lett. {\bf 105},  140501  (2010).

\bibitem{Mohammady2010}
M.~H. Mohammady, G.~W. Morley, A. Nazir, and T.~S. Monteiro, Phys. Rev. B {\bf
  85},  094404  (2012).

\bibitem{Mohammady2010a}
M. Mohammady, G.~W. Morley, and T.~S. Monteiro, Phys. Rev. Lett. {\bf 105},
  067602  (2010).

\bibitem{Balian2012}
S. Balian, M. Kunze, and M. Mohammady, Arxiv preprint:1203.5607, Accepted in
  Phys. Rev. B  (2012).

\bibitem{Riemann2006}
H. Riemann, N. Abrosimov, and N. Noetzel, ECS Transactions {\bf 3},  53
  (2006).

\bibitem{Becker2010}
P. Becker, H.-J. Pohl, H. Riemann, and N. Abrosimov, Physica Status Solidi (a)
  {\bf 207},  49  (2010).

\bibitem{Stoll2006}
S. Stoll and A. Schweiger, J. Mag. Res. {\bf 178},  42  (2006).

\bibitem{Schweiger2001}
A. Schweiger and G. Jeschke, {\em {Principles of pulse electron paramagnetic
  resonance}} (Oxford University Press, Oxford, 2001), p.\ 578.

\bibitem{Feher1959a}
G. Feher and E. Gere, Phys. Rev. {\bf 114},  1245  (1959).

\bibitem{DeSousa2003a}
R. de~Sousa and S. {Das Sarma}, Phys. Rev. B {\bf 67},  033301  (2003).

\bibitem{Jr1963}
T. Castner, Phys. Rev. {\bf 130},  58  (1963).

\end{thebibliography}

\end{document}